\documentclass[twocolumn, times]{aastex631}

\usepackage{amsmath, latexsym, color, verbatim}

\shorttitle{IFU Adaptive Trace Modeling}
\shortauthors{Law \& Clarke 2026}

\newcommand{\Ha}{\ensuremath{\rm H\alpha}}
\newcommand{\Hb}{\ensuremath{\rm H\beta}}

\newcommand{\othree}{\textrm{[O\,{\sc iii}]}}

\begin{document}

\title{Mitigating Resampling Artifacts for the JWST IFU Spectrometers with Adaptive Trace Modeling}

\author[0000-0002-9402-186X]{David R.\ Law}
\affiliation{Space Telescope Science Institute, 3700 San Martin Drive, Baltimore, MD, 21218, USA}

\author[0009-0002-3561-4347]{Melanie\ Clarke}
\affiliation{Space Telescope Science Institute, 3700 San Martin Drive, Baltimore, MD, 21218, USA}

\begin{abstract}

The integral-field unit (IFU) spectrometers on board the James Webb Space Telescope (JWST) undersample the nearly diffraction-limited point spread function provided by the telescope optics.  This undersampling produces large oscillating spectral artifacts when the data is resampled into regularly-gridded data cubes, which poses a significant challenge for many scientific analyses.  We describe here a generalized technique to use cubic basis spline models to interpolate the observed spectral traces onto a higher-resolution grid prior to data cube rectification, which largely eliminates these artifacts in addition to helping reduce biases in point source spectra from clusters of bad pixels.  We demonstrate the utility of this adaptive resampling technique for a variety of JWST NIRSpec and MIRI MRS observations ranging from isolated point sources to embedded AGN, crowded stellar fields with diffuse emission, and protostars with rich molecular bands.

\end{abstract}



\section{Introduction}
\label{intro.sec}

According to the 
Nyquist-Shannon theorem, the sampling frequency of any signal must be at least twice as fast as the highest-frequency component of that signal in order to avoid aliasing power into lower frequency components during any resampling of the original signal.  One of the direct implications of this for standard astronomical observations is that
we must sample the point spread function (PSF) with at least two detector pixels per FWHM in order to avoid introducing artifacts when resampling data from the native pixel grid as is commonly done when combining dithered observations.  This requirement can be particularly challenging for space-based instruments that achieve near diffraction-limited performance, and the effects of undersampling have been well documented for the Hubble, Spitzer, and James Webb space telescopes.

\citet{anderson16} for instance noted the impact of undersampling on reconstruction of the point-spread function for HST/WFC3 imaging data, which can complicate efforts to perform outlier detection using resampled data.  In the spectral domain artifacts can be even more obvious, as spectra dispersed on a grid of detector pixels are typically tilted with respect to that grid, resulting in pixel samplings that differ as a function of wavelength.
\citet{dressel07} noted this effect for HST/STIS, while \citet{smith07} described its impacts on efforts to create spectral maps from the Spitzer telescope's Infrared Spectrograph.  More recently, \citet{law23} described how undersampling can produce resampling artifacts in spectral data cubes for the slicer-type 
NIRSpec \citep{boker22} and MIRI \citep{wright23} integral-field units aboard the James Webb Space Telescope.

As demonstrated by \citet{law23}, these resampling artifacts manifest as low-frequency oscillations in the spectra of drizzled data cubes, with the oscillation frequency being wavelength-dependent and tied to the tilt angle of the dispersed trace with respect to the detector pixel grid.
These artifacts are strongest in the spectra of individual cube spaxels near unresolved point sources, are partially mitigated by dithering, and average out when extracting spectra from apertures comparable in size to the observational PSF.

Immediate solutions to the problem are thus to either perform scientific analyses in the native detector pixel space, or to extract spectra from resampled data cubes using sufficiently large extraction apertures.  
Neither solution is ideal.  Rectified and coadded data cubes, for instance, are much more convenient and easy to work with than the often-complex formats of the native detector-level data.  At the same time, crowded astronomical scenes often require analysis of extremely small areas of sky in order to minimize source confusion.  Indeed, a common method of IFU data analysis is to apply spectral fitting techniques to the spectra of each cube spaxel, and thus build up 2d maps of line emission, kinematics, or a host of other physical properties \citep[e.g.,][]{westfall19}.

Various groups have thus developed techniques to correct for these artifacts in JWST data.
As part of the GA-NIFS survey for instance, \citet{perna23} corrected for resampling artifacts via sinusoidal modeling of spectra from the resampled cubes.  \citet{dumont25} likewise developed the {\sc wicked} package using Fourier analysis to model and remove regular sinusoidal features from affected cube spaxels.  Most recently, \citet{shajib25} developed the software package {\sc raccoon} using sinusoidal chirp functions to correct the data cubes with template spectra derived from small apertures.
All three implementations can be effective for specific science cases, but intrinsically rely on measuring and removing the artifacts after they have already been introduced into the data.
As such, all require extensive customization to ensure that genuine spectral features do not bias the correction routines.

Here, we present a turn-key method that uses adaptive PSF-modeling of spectral traces in the native detector space to avoid creating such artifacts in the IFU data cubes in the first place, thereby allowing it to handle arbitrary scenes and source spectra with little to no tuning.
We describe our algorithm in \S \ref{method.sec}, and evaluate its performance for point sources observed with both the MIRI MRS and NIRSpec IFU in \S \ref{performance.sec}.  We discuss the application of this technique to a variety of astronomical data sets in \S \ref{discussion.sec} and Appendix \ref{appendix.sec}.  We summarize our results in \S \ref{summary.sec}.



\section{Adaptive Trace Model (ATM) Method}
\label{method.sec}

As outlined in \S \ref{intro.sec}, the general problem of mitigating artifacts that arise when resampling undersampled data is unsolvable as the necessary information is irretrievably lost.
However, in our specific case it is possible to bring to bear additional information provided by our knowledge of astronomical optics and the physical properties of the instruments.
In particular, for well-calibrated instruments we know that the PSF should change only slowly with wavelength, and that unresolved point sources should have centroids that remain fixed in celestial coordinates at all wavelengths.
These two assumptions allow us to take advantage of the 
very pixel boundary crossings that produce the sinusoidal artifacts in the first place, and are sufficiently powerful that they can largely eliminate the resampling problem in most JWST observations.

\begin{figure*}[!htbp]
\epsscale{1.2}
\plotone{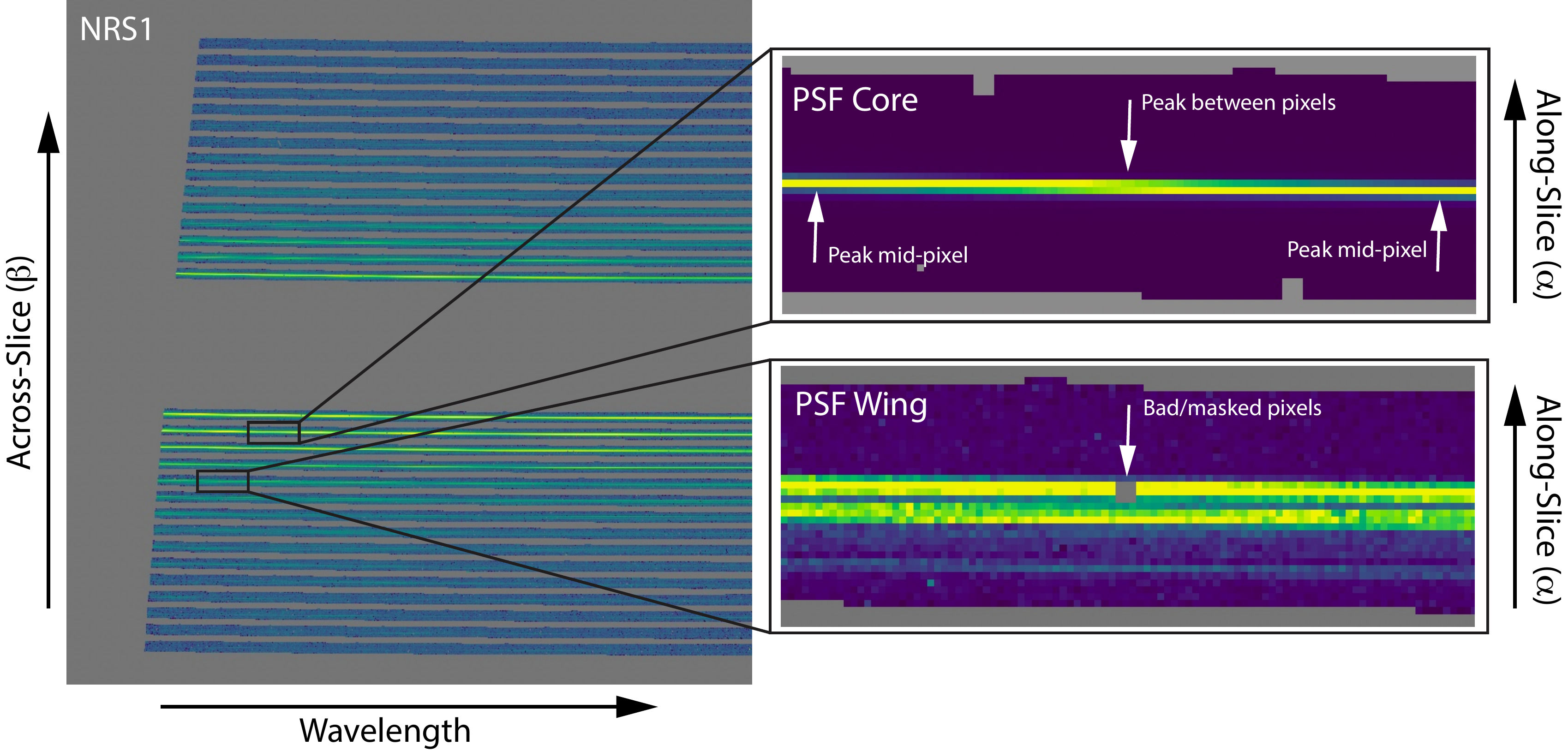}
\caption{Left panel: NIRSpec NRS1 calibrated detector image of spectrophotometric standard star J1757132 (spectral type A8Vm) in the G395H/F290LP spectral configuration.  Image uses a logarithmic stretch to show the dispersed trace of the star illuminating multiple of the 30 separate IFU slices.  Right panel: Zoom-in on regions of the PSF core and wings illustrating the change in pixel-phase sampling of the PSF as a function of wavelength.  Zoomed panels use a variable linear stretch to accentuate details.  The across-slice coordinate $\beta$ changes between slices, while the along-slice coordinate $\alpha$ changes within each slice.
}
\label{detector.fig}
\end{figure*}

In Figure \ref{detector.fig} we show a NIRSpec IFU observation of an unresolved point source obtained by JWST program ID (PID) 3399 (PI: M. Perrin).  In the left-hand panel we show the calibrated (`$\ast\_$cal.fits') image from the NRS1 detector.  Each of the horizontal stripes corresponds to one of the 30 IFU slices, arranged in the order of their across-slice position $\beta$.  The spectrum of each slice is dispersed horizontally with increasing wavelength to the right; within a given slice each column thus corresponds to a different wavelength.
While the FWHM of this point source is comparable to the width of a single slice, the star is bright enough that the wings of the PSF illuminate multiple slices out to a distance of about 0.5 arcsec.

Zooming in on the dispersed trace of the bright PSF core (Figure \ref{detector.fig}, right-hand panel) we show that each stripe is about 30 pixels tall in the along-slice spatial direction $\alpha$.  Each slice can thus be thought of as a dispersed long-slit spectrum; where a slice covers the PSF core the normalized spatial profile is narrow, and where a slice covers the PSF wings we observe Airy rings and other structures characteristic of the JWST optics.
The detector pixels undersample the PSF, and thus the
illumination pattern for the core alternates between a single bright pixel surrounded by two faint neighbors, to two adjacent pixels of comparable brightness, and back again.  This is caused by the small tilt of the spectral trace with respect to the detector pixel grid; the PSF is sampled in a different pixel `phase' at different wavelengths.

\begin{figure}[!htbp]
\epsscale{1.2}
\plotone{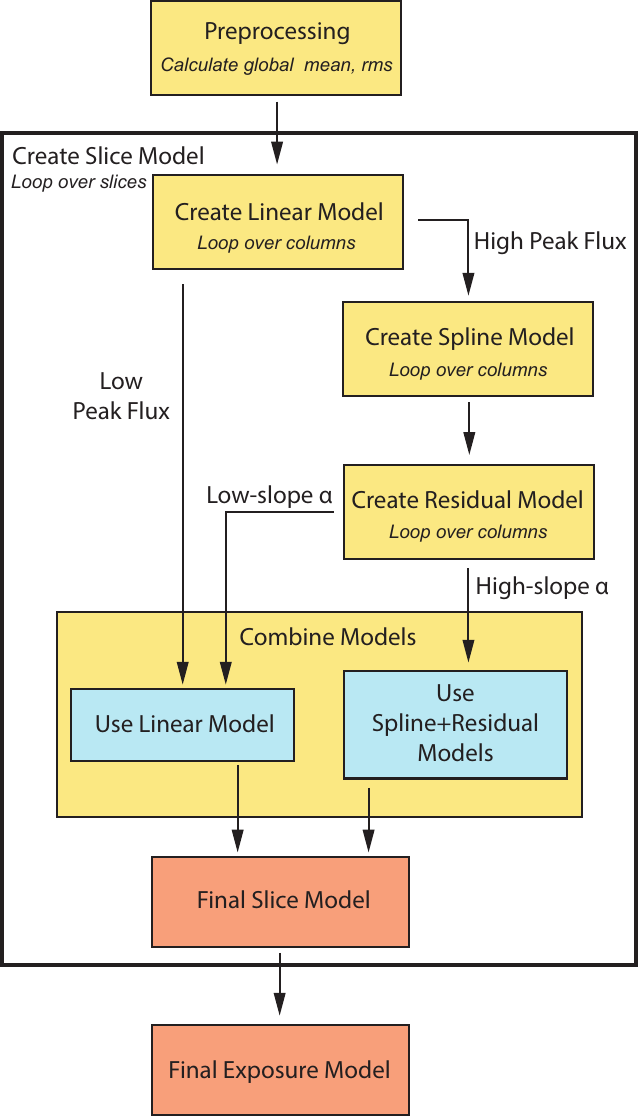}
\caption{Flowchart outlining the adaptive trace model method for oversampling a given exposure.  Each IFU slice is treated
independently.  Spline models are created only for slices with significant emission, and used only for along-slice locations with steep gradients indicative of undersampling.
}
\label{flowchart.fig}
\end{figure}

We define the original calibrated detector image  to be $f_{\rm orig}$.  In the case of NIRSpec, $f_{\rm orig}$ is a 2048$\times$2048 pixel array in which all pixels within spectral slices are flux-calibrated in units of MJy/sr, and all pixels that do not see light from the sky have values of NaN (not-a-number).
Our aim is to construct a model for these data at each wavelength, and to use this model to resample each $f_{\rm orig}$ onto a higher-resolution image $f_{\rm interp}$ with $N$ times the spatial sampling in the along-slice direction.  NIRSpec disperses data horizontally and will thus map a 2048$\times$2048 pixel array to a 2048$\times$4096 pixel array for $N=2$, while MIRI MRS disperses data vertically and will map a 1032$\times$1024 pixel array to a 2064 $\times$ 1024 pixel array for $N=2$.  For clarity we will focus our discussion on the NIRSpec case, but note that MIRI will follow a similar approach modulo a 90$^{\circ}$ rotation.  To do so, we 
define three intermediate products $f_{\rm linear}$, $f_{\rm spline}$, and $f_{\rm resid}$ as described in \S \ref{linear.sec} --- \ref{residual.sec}, and combine them as described in \S \ref{combined.sec}.
A flowchart providing an overview of our approach is shown in Figure \ref{flowchart.fig}.

\subsection{Preprocessing}
\label{preproc.sec}

We first preprocess $f_{\rm orig}$ by computing the sigma-clipped mean and rms of the entire image.
The rms is saved for later use in \S \ref{spline.sec}, and the mean is evaluated to determine whether it is greater than or equal to zero.  In almost all cases this will be true and no correction is necessary.  In the rare cases in which the mean is negative (typically due to calibration errors such as oversubtraction of the dark current) we subtract the mean from the detector image prior to further processing to ensure that all of our models will have a baseline around zero with some scatter due to observational uncertainty.
Since the MIRI MRS has four wavelength channels that disperse simultaneously onto two detectors, these statistics must be calculated for each channel separately.

Likewise, we define a slice mask identifying the pixels associated with each IFU slice.  The models for each slice will be determined independently in \S \ref{linear.sec} --- \ref{combined.sec}.

\subsection{Linear Model}
\label{linear.sec}

The first model that we construct is a simple linearly-interpolated model $f_{\rm linear}$.
This interpolation is accomplished column by column (i.e., wavelength by wavelength) in each slice.  In each column we identify the rows that contain flux and map these to the output grid with a factor of $N$ oversampling.
In the case of $N=2$ for instance, pixel fluxes corresponding to column 358 and rows [98, 99, ..., 127, 128] in the original image will be interpolated to the fractional pixel positions [97.75, 98.25, 98.75, ..., 127.75, 128.25] and the result inserted into column 358 and rows [196, 197, ... 256, 257] in the output grid.\footnote{In practice, it makes little to no difference whether these fractional pixel positions are slightly shifted or include the original sample points (e.g., [97.75, 98.25, 98.75, ...] vs [98.0, 98.5, 99.0, ...]).}
Such a linear interpolation is similar to what is being performed by the 3d drizzle algorithm \citep{law23} during data cube construction, and is fully sufficient to describe the intensity distribution everywhere that the input flux distribution is varying smoothly.  Indeed, even near unresolved point sources this linear interpolation alone can slightly reduce resampling noise as it ensures that no one input detector pixel dominates the resampled value for a given output cube voxel.

\subsection{Spline Model}
\label{spline.sec}

The second model that we construct is a cubic basis spline model that is similarly built column by column across each slice.  We start with the columns for which the angle between the dispersed trace and the detector pixel grid is the largest (i.e., different pixels will have the largest sampling phase difference), and work our way toward the other side of the detector.  This thus corresponds to starting at the left-hand edge of the NRS1 detector and the right-hand edge of the NRS2 detector.  In the case of the MIRI MRS we start at the top and bottom of the detector, and work our way into the middle where the traces are parallel to the pixel grid.

The $\sim$ 30 pixels in each column provide only a limited number of samples with which to model the along-slice profile, and we therefore 
define a range of adjacent columns (i.e., wavelengths) whose data we will use to assist in constructing the 1d spline model for the column.  This range should ideally be as small as possible while still ensuring sufficient phase coverage to adequately reconstruct the source profile; in practice we find 50 adjacent columns in each direction to be sufficient.  We define a normalized image $f_{\rm norm}$ in which the pixel values in each column are divided by the sum of values in that column, thereby normalizing out the intrinsic spectrum of the source.\footnote{Given the redundancy across many columns, this normalization is typically sufficient even in the presence of a modest number of bad and/or missing pixel values.}

\begin{figure*}[!htbp]
\epsscale{1.2}
\plotone{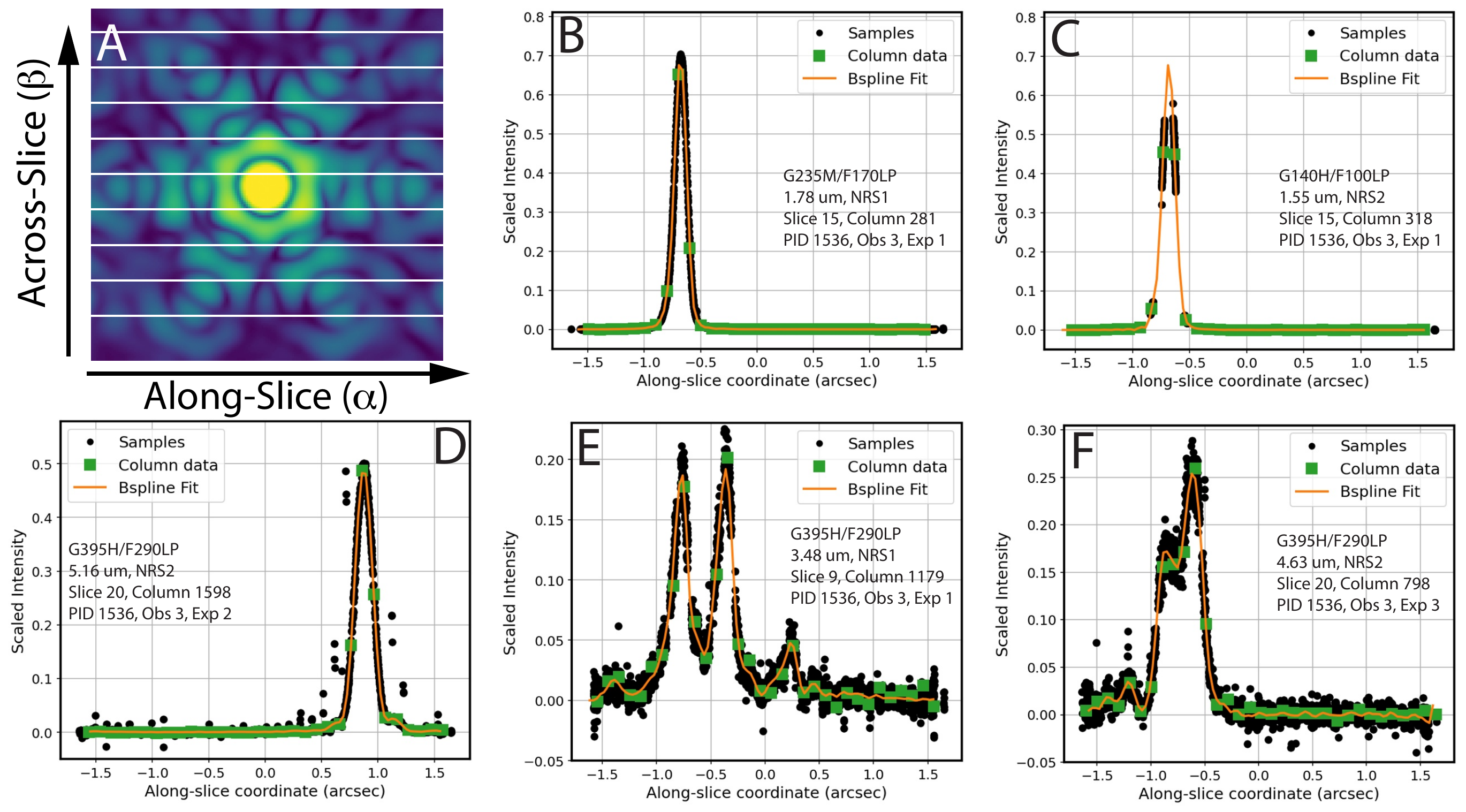}
\caption{Panel A: Simulated JWST 2$\micron$ PSF illustrating the presence of Airy rings and other complex features due to the telescope optics, with illustrative slice boundaries shown in white.  Panels B-F: Real detector profiles as a function of the along-slice coordinate $\alpha$ for observations of standard star J1743045 in various NIRSpec grating settings, slices, and wavelengths.  Solid green squares represent the calibrated detector data in the indicated column, solid black points represent data from the neighboring $\pm$50 columns normalized to a common summed value.  The orange line represents the derived basis-spline fit to the black points.
}
\label{bspline.fig}
\end{figure*}

We plot the normalized flux from each pixel in this range of columns as a function of their along-slice coordinate $\alpha$ in Figure \ref{bspline.fig}.\footnote{In practice $\alpha$ is measured in physical units of millimeters for NIRSpec; for illustrative purposes in this manuscript we use the approximate conversion factor of 0.25 arcsec mm$^{-1}$.}.  Since the PSF changes only slowly with wavelength 
the normalized data from the adjacent columns gives many measurements of the along-slice profile with different pixel phase samplings.  As illustrated by Figure \ref{bspline.fig}, this gives an excellent sampling of the actual 1d profile shape (black points) that is far superior to that achieved using pixels from the central column alone (green squares).

Following well-established techniques in the astronomical literature \citep[see, e.g.,][]{bolton07, law16}
we fit the profile defined by these black points with a 1d cubic basis spline function .  We iteratively reject points more than 2.5$\sigma$ away from the spline fit, which takes care of any improper normalizations due to noise or bad pixels in any individual column.  
Significant care must be taken in selecting the
number and spacing of the basis spline knots to ensure that the spline is fully able to describe all real structures in the profile, but not overfit artifacts.  In general, we want approximately twice the number of spline knots as there are independent detector rows in a given slice.  Empirically, we find that 68 and 36 knots distributed uniformly across the slice works well for the NIRSpec IFU and MIRI MRS respectively.\footnote{Since the NIRSpec and MIRI IFUs have slice widths of approximately 30 and 15 pixels respectively, by using $\pm 50$ adjacent columns in constructing the spline model there are thus approximately 50 data points between each knot.}

As indicated by Figure \ref{bspline.fig}, the normalized along-slice profile can have significant and complex structure, even in the case of a simple point source.  Although the profile is a smooth, single peak for slices that sample the center of the PSF (panel B), complex and broad multi-peaked profiles (panels E and F) are common away from the PSF center where Airy rings and diffraction spikes dominate \citep[see, e.g.,][]{perrin12}.  In all of these cases the basis spline model does a good job of reproducing the observed profile, although at lower SNR (panels E and F) the scatter becomes larger.

This spline model is easy to construct in regions of the detector where the dispersed trace crosses pixel boundaries often.  On the left edge of NRS1 for instance, the trace completes a full phase of sampling from centered-in-pixel to centered-in-pixel again every 80 columns.  At the right hand edge of NRS1 however, the trace angle flattens until it is nearly parallel to the pixel grid.  This means that in a given window of adjacent columns there may no longer be full representation of all pixel phases with which to constrain the spline model (e.g., Figure \ref{bspline.fig} panel C).  We find that the spline-fitting technique is robust against this loss of information so long as the gaps between samples are smaller than $\sim$ 1.6 times the knot spacing.  When the gaps become larger, there is insufficient information to fully constrain the model and it can start to fit spurious features.

This can be mitigated by increasing the range of adjacent columns used to construct the profile, albeit at the cost of a progressively inaccurate model.  Instead, we opt to simply save the spline profile fit from each column, and use the saved fit from the previous column when the sampling in any column exceeds 1.6 times the knot spacing.  Such extrapolation is necessary for the right-most $\sim 200$ pixels of the NIRSpec NRS1 detector, a similarly-sized region at the left-hand edge of NRS2, and the central 30-300 pixels of the MIRI MRS.\footnote{The width of this extrapolation region is highly channel-dependent according to the intrinsic sampling and trace curvature in each channel.  Typical values are 300, 100, 60, and 30 pixels in Channels 1-4 respectively.  Since our modeling for MIRI starts at both ends of the spectrum however, the actual distance over which any given interpolation is used is roughly half this number of pixels.}  In the simple case of a single star we find that the impact of this extrapolation 
on the voxel fluxes in the final
reconstructed data cubes is about 0.9\% (1$\sigma$ rms) overall in the affected regions, and $\sim 0.2$\% for the brightest spaxels.

Once we have obtained the normalized spline model for a given column we evaluate it at the original Y pixel coordinates and determine the necessary scaling factor to reproduce the observed values $f_{\rm orig}$.  We determine this from 
the flux-weighted mean ratio between the original pixel fluxes and the spline model, in which weights are given by the model flux and set to zero anywhere the data or model was negative or NaN-valued.
The sigma-clipped mean and rms of the ratios of the 5 largest-weight points are computed and used to reject any points that differ at more than 2$\sigma$ to help reject individual bad pixels in the fit.  The remaining weights are normalized and used to compute the weighted mean scale factor, which is robust against a small number of missing and/or saturated pixels.
Finally, the scaled basis spline model is evaluated at the oversampled fractional pixel positions (see \S \ref{linear.sec}) to obtain the corresponding values for $f_{\rm spline}$.

In practice, we note that it is often unnecessary to apply the spline modeling to a given slice if that slice contains no significant source flux.  We therefore skip this step if the maximum flux in a slice is less than $10\sigma$ above our previous estimate of the sigma-clipped detector averaged count rate.

\subsection{Residual Model}
\label{residual.sec}

\begin{figure}[!htbp]
\epsscale{1.2}
\plotone{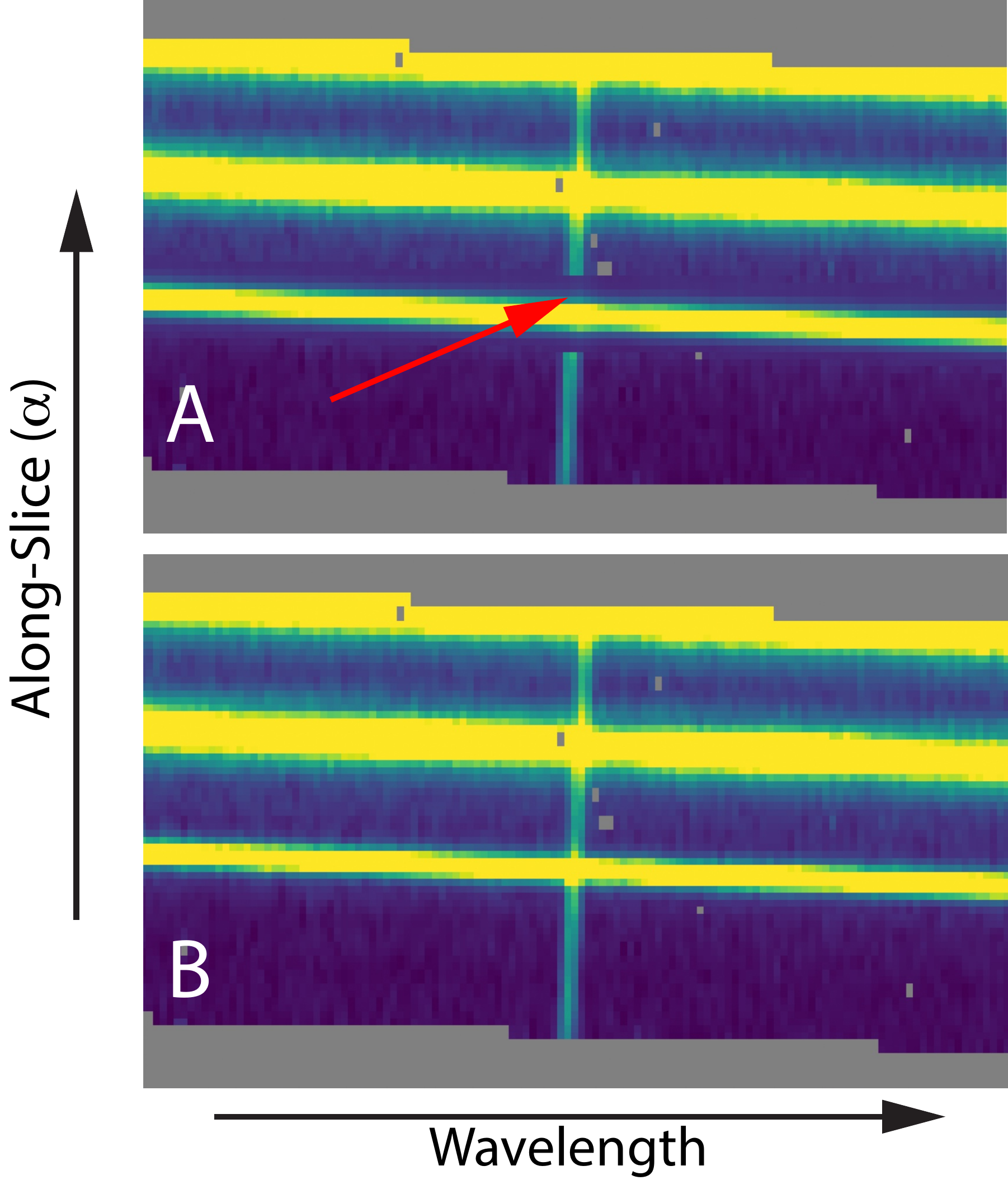}
\caption{Top panel: Zoom-in of a NIRSpec oversampled detector image showing missing full-field line emission (narrow vertical features) in the vicinity of the lower stellar spectral trace (horizontal features) that has been oversampled using the basis-spline technique.  In contrast, the upper traces oversampled using the linear technique show no such missing emission.  Bottom panel: Same oversampled image after application of the residual correction described in \S \ref{residual.sec}.
}
\label{complex2.fig}
\end{figure}

By construction, the spline-modeling technique assumes that the source profile varies only slowly with wavelength.  This assumption is reasonable for isolated point sources, but is frequently invalid for other common astrophysical targets.  In the case of stars embedded in a nebula, for instance, there may be narrow emission line features that fill the field of view at discrete wavelengths.  As illustrated in Figure \ref{complex2.fig} (top panel), the spline model can artificially suppress such features since they are deviations from the usual spatial profile.

We therefore also construct a residual correction model to account for any structure that does not conform to the spline model expectations.  We do so by evaluating the spline model at the original detector pixel positions, and subtracting it from the original data $f_{\rm orig}$ to obtain a residual image.  This residual image is then oversampled using linear interpolation similar to that described in \S \ref{linear.sec} in order to obtain the correction image $f_{\rm resid}$.
While this residual correction reduces the SNR gains that might be provided by the spline model alone, it corrects for any biases introduced by the model.  This fixes the missing emission lines shown in Figure \ref{complex2.fig}, and can similarly account for other structure not well captured by a slowly-varying spline model (e.g., two nearby point sources with different spectral features).


\subsection{Constructing the Combined Model}
\label{combined.sec}

Once the loop over all columns within a given slice has been completed, we then combine $f_{\rm linear}$, $f_{\rm spline}$, and $f_{\rm resid}$ together to create the final adaptive trace model (ATM) result $f_{\rm interp}$.  Each component has its own strengths and weaknesses; $f_{\rm linear}$ is unbiased but does not increase the effective resolution, $f_{\rm spline}$ can increase the effective resolution at the cost of computational runtime (see \S \ref{pipeline.sec}) and systematic biases in the vicinity of unmodeled structures, and $f_{\rm resid}$ can mitigate systematic biases at the cost of increased random noise. 

Using the normalized spline models such as those shown in Figure \ref{bspline.fig}, we differentiate the model fits as a function of the along-slice position $\alpha$.  Rows where the slope is greater than some threshold value (taken here to be 0.1 pixel$^{-1}$) identify those $\alpha$ where the model is changing the fastest and is thus the least well sampled by the native detector pixels.
All rows near these high-slope regions are the highest priority regions for using the spline model.  We find that a `padding' range of about $\pm$ 2 -- 3 native detector pixels around these high-slope regions suffices to identify this crucial range well.\footnote{It is important to ensure that any such identified high-slope regions are defined uniformly to all wavelengths within a given slice to avoid introducing artifacts where the linear and spline models join.}

\begin{figure*}[!htbp]
\epsscale{1.2}
\plotone{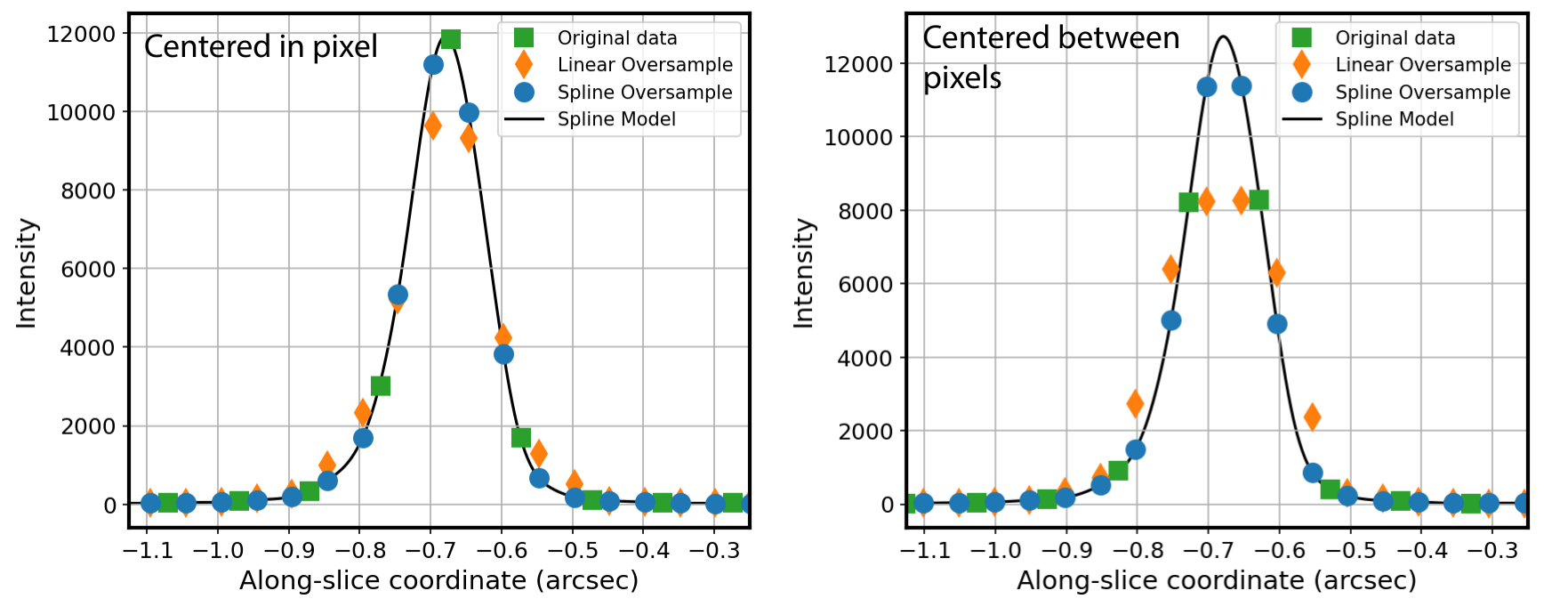}
\caption{Oversampled slice profiles for cases where the PSF peak is centered in a pixel (left panel) vs between pixels (right panel).  Green squares represent the original data in a given column; orange points show the oversampled values from simple linear interpolation between the original points.  Blue points show the $N=2$ oversampled values from the spline model fit (black curve).
}
\label{extrapolate.fig}
\end{figure*}

The final oversampled image $f_{\rm interp}$ is thus defined as the sum of $f_{\rm spline}$ and $f_{\rm resid}$ in these high-slope regions and $f_{\rm linear}$ everywhere else, providing a result that is fully data-driven and independent of any specific PSF model.  This oversampling is particularly critical in regions where the PSF peak was centered between two detector pixels.  As illustrated by Figure \ref{extrapolate.fig} (right-hand panel), the spline fitting technique does a good job of recovering the true peak of the emission profile when such information would be lost in cubes resampled from either the original data or simple linear interpolation methods.

The amount of data interpolated using the spline vs linear techniques will depend on the chosen flux limit, slope limit, and pixel padding range.  In practice, for a typical NIRSpec point source and the default parameters given here spline models will be computed for roughly 5 slices, allowing the remaining 25 slices to be more rapidly oversampled with the linear technique.  In those 5 slices, roughly 15-30\% of the detector pixels will actually be interpolated using the spline model, focusing on the central regions of the spectral trace where the spline model advantages are greatest.
These defaults will need to be adjusted for certain science cases; crowded fields in particular will need to ensure that the spline plus residual model is used everywhere to ensure proper modeling of both bright and faint sources.

\section{Performance}
\label{performance.sec}

The world coordinate solution (WCS) mapping each pixel in the ATM result $f_{\rm interp}$ to its corresponding wavelength and position on the sky can be trivially obtained by evaluating the WCS of the original data at the relevant sub-pixel positions.  Rectified data cubes can thus be constructed using the oversampled pixel inputs for each dithered exposure following the standard 3D drizzle method described by \citet{law23}.  In \S \ref{nrsperf.sec} and \ref{mrsperf.sec} we evaluate the impact of the ATM technique on point source spectra extracted from such oversampled data cubes for the NIRSpec and MIRI IFUs, and assess the computational performance impact on the JWST pipeline in \S \ref{pipeline.sec}

\subsection{NIRSpec IFU}
\label{nrsperf.sec}

\begin{figure*}[!htbp]
\epsscale{1.2}
\plotone{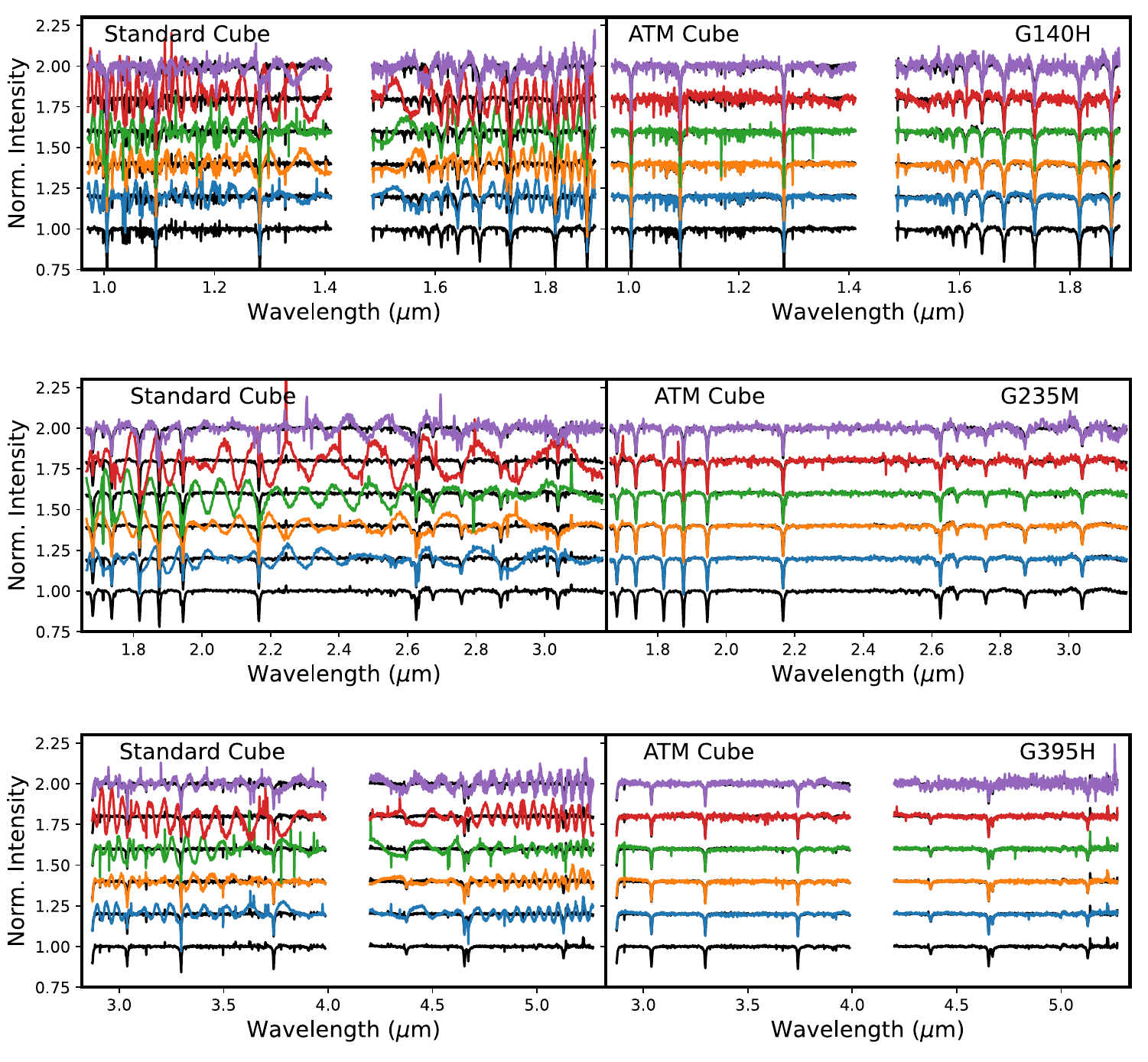}
\caption{Resampling noise in standard (left-hand column) and $N=3$ adaptive trace model (ATM) oversampled data cubes (right-hand column) for single-spaxel spectra extracted from observations of standard star J1743045 (spectral type A5 IIIm).  Top, middle, and bottom rows respectively show results for observing configurations G140H/F100LP, G235M/F170LP, and G395H/F290LP.  The solid black lines represent the aperture-integrated spectra, while colored lines represent single-spaxel spectra extracted from the first, second, 4th, 8th, and 16th brightest spaxels (blue through purple respectively).  Aperture-integrated spectra have been underplotted beneath all single-spaxel spectra to aid comparisons.  All spectra have been normalized by a low-order polynomial fit to the spectrum and arbitrarily offset vertically for clarity.  Narrow features due to hydrogen absorption in the stellar spectrum are most noticeable in composite (black lines) and ATM-corrected spectra.
}
\label{results.fig}
\end{figure*}

We test the performance of the ATM technique for the NIRSpec IFU using observations of standard star J1743045 provided by the Cycle 1 flux calibration program PID 1536 (PI: K. Gordon).  This program observed this A-type star in each of the NIRSpec filter/grating settings using a standard 4-point dither pattern, and the JWST calibration pipeline produces data cubes with a default scale of 0.1 arcsec spaxel$^{-1}$.

In Figure \ref{results.fig} we show the normalized spectra of an assortment of the brightest individual spaxels in both the standard and ATM-oversampled data cubes.  Resampling artifacts in these single-spaxel spectra are obvious in the standard cubes 
and have average half-amplitudes of 9\%, 8\%, and 6\% respectively in the brightest 20 spaxels for G140H, G235M, and G395H.  Using our ATM technique with an oversampling factor $N = 2$, these amplitudes decrease to 3\%, 2\%, and 2\% respectively.  Increasing the oversampling to $N = 3$, the amplitudes all decrease to the noise floor $\sim 1$\% with few visible artifacts remaining (Figure \ref{results.fig}, right-hand panels) and in which it is easy to see the intrinsic hydrogen absorption features in the stellar spectrum.  Additional oversampling by factors of four or five confers no clear additional gains, either visually or in the computed residual amplitudes.

At the same time, the summed spectra within a 0.6 $\times 0.6$ arcsec box show excellent flux conservation
with the original and spline-oversampled spectra consistent as a function of wavelength to within $0.1 \pm 0.3$\% for G140H and G235M, and to within $0.2 \pm 0.1$\% for G395H.  Deviations between the summed spectra of larger than 1\% are typically due to artifacts narrower than the spectral resolution in the original cube that are not present in the corrected data.

\begin{figure*}[!htbp]
\epsscale{1.2}
\plotone{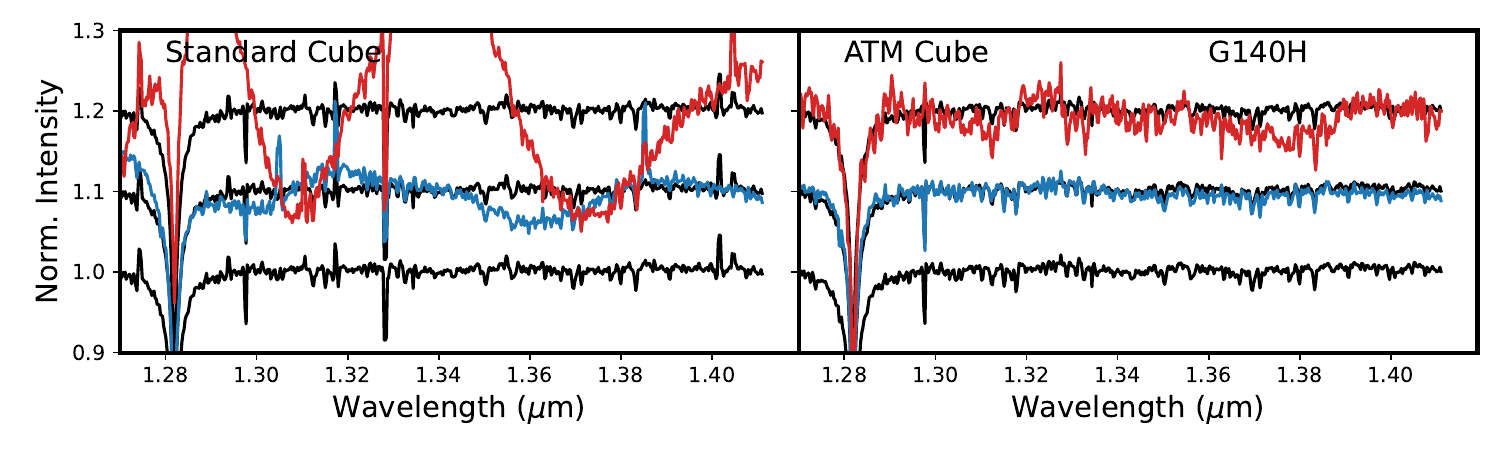}
\caption{As Figure \ref{results.fig} but highlighting the limitations of the method by
zooming in on a small wavelength range for the 1st and 8th-brightest spaxels
(blue and red lines respectively) for G140H.
}
\label{g140hresid.fig}
\end{figure*}

Despite the clear improvements in the individual spaxel spectra compared to the standard data cubes, we caution that some sampling artifacts can remain, albeit at substantially lower amplitudes.  In Figure \ref{g140hresid.fig} we zoom in on one of the spectra from the G140H data cube that showed the largest deviations from the aperture-summed spectrum.
While the brightest spaxel is consistent with the aperture-summed spectrum (black line) in the displayed wavelength range to 
an RMS of 0.6\% in the ATM-corrected cube (compared to 2\% in the standard cube), the 8th-brightest spaxel deviates at 1.8\% RMS (compared to 11\% in the standard cube) and has an artificial dip around 1.38 $\micron$.  Typically this is due to convergence issues in the spline model as the wings of the PSF can be biased by bad pixels.  The ATM-corrected per-spaxel spectra should thus be treated with a degree of caution at the percent level; when in doubt we note that genuine spectral features should be present in data cubes made from each of the invididual exposures.


\subsection{MIRI MRS}
\label{mrsperf.sec}

Resampling noise is much more complex for the MIRI MRS, as the sinusoidal patterns produced by resampling beat against the intrinsic fringing produced by constructive and destructive interference within the MIRI instrument \citep[see, e.g.,][]{argyriou20}.  The nature of this intrinsic fringing is dependent upon the size, brightness, and location of objects within a given scene, and a 1d residual fringe correction is thus typically applied to MRS spectra to remove Fourier modes corresponding to known instrumental fringe frequencies.

\begin{figure*}[!htbp]
\epsscale{1.1}
\plotone{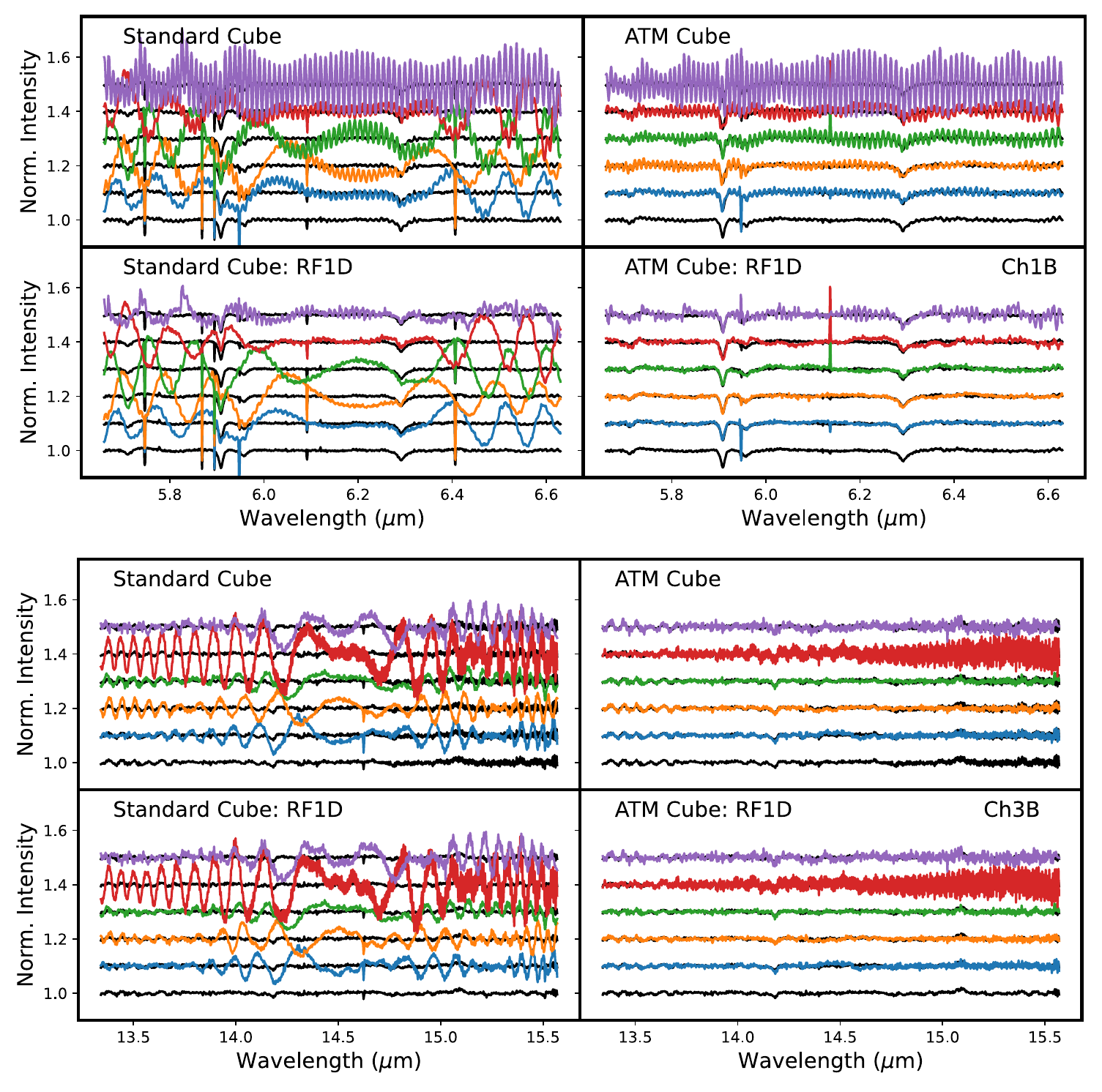}
\caption{As Figure \ref{results.fig} but for MIRI MRS Ch1A and Ch3B observations of bright standard star HD2811 (spectral type A3 V).  In each case we show the extracted spectra with and without correction for high-frequency fringing using the pipeline 1d residual fringe correction routine (RF1D).
}
\label{miri.fig}
\end{figure*}

We illustrate the typical characteristics of resampling noise in MRS data in Figure \ref{miri.fig} using 4-point dithered observations of A-type standard star HD 2811 from PID 1536.  Individual-spaxel spectra show both high-frequency artifacts due to fringing and low/variable frequency artifacts due to resampling noise \citep[see also][their Figure 9]{law23}.  These resampling artifacts have typical half-amplitudes of about 10\% in Channel 1A and 7\% in Channel 3B, and are well corrected by the spline oversampling algorithm.
Using an oversampling factor $N = 2$ the amplitudes decrease to 3\% and 2\% respectively, decreasing to 1\% with an oversampling factor $N = 3$ with few obvious artifacts remaining (as can be seen from the nearly-unity ratios between single-spaxel and summed spectra in Figure \ref{miri.fig}, right-hand panels).
Similarly to NIRSpec, the absolute flux of the integrated spectrum is conserved to $0.3 \pm 0.2$\% for Ch1A and $0.2 \pm 0.1$\% for Ch3C.

\subsection{JWST Pipeline Performance}
\label{pipeline.sec}

This technique has been implemented as an optional step \textsc{adaptive\_trace\_model} in build 12.3 of the JWST calibration pipeline \citep{bushouse25} scheduled for release in April 2026.  In most cases (e.g., those presented in \S \ref{performance.sec}, \S \ref{discussion.sec}, and three of four cases in the Appendix \ref{appendix.sec}) no user adjustment of the default parameters is necessary.  The limiting flux threshold above which to derive spline models, and the slope threshold at which to use those models are exposed as customizable parameters though, which can be useful for extremely rich and complex scenes.

Using a standard Mac laptop with 64 GB of RAM, the runtime of our pure-python implementation of the ATM technique for the NIRSpec IFU is about 12 seconds per slice for which a spline model is created.\footnote{MIRI has a factor of about 4 fewer pixels, and thus is about 4x faster on a per-exposure basis.}  A simple scene therefore takes about 90 seconds per exposure, while complex scenes that create spline models for all 30 slices can take up to 6 minutes per exposure.

We note that use of the ATM method also imposes additional runtime overheads on the {\sc pixel\_replace} and {\sc cube\_build} pipeline steps.  The {\sc pixel\_replace} step attempts to fill in NaN-valued pixels via linear interpolation that were not already replaced by spline-derived values in the ATM step.  Since there are roughly 3x as many such pixels after oversampling by a factor of $N = 3$ the runtime of this step approximately triples from 
30 seconds to 90 seconds for a typical exposure.
Likewise, for $N = 3$ the {\sc cube\_build} step that resamples the native detector pixels into a rectified 3D data cube also has a 3x larger list of input pixels to process.  In this case the architecture of the step is such that the runtime approximately doubles, increasing from 50 seconds to about 100 seconds for an association of four exposures using the NRS1 detector.


\section{Discussion}
\label{discussion.sec}

The utility of the ATM algorithm lies not in its application to isolated point sources however, but rather to cases where such point sources are embedded within extended structure that we wish to disentangle from resampling artifacts.
We therefore test the algorithm on a typical science case in which a high-redshift QSO is embedded within a host galaxy and driving outflows traced by ionized gas emission.  We select the QSO LBQS 0302-0019 from JWST PID 1220 (PI: N. Lutzgendorf); this system was studied extensively by \citet{perna23}, who used these data to develop their own method of correcting resampling noise via sinusoidal modeling.
Given the results of \S \ref{performance.sec}, we choose to oversample this and all future test cases by a factor of $N = 3$.

\begin{figure*}[!htbp]
\epsscale{1.2}
\plotone{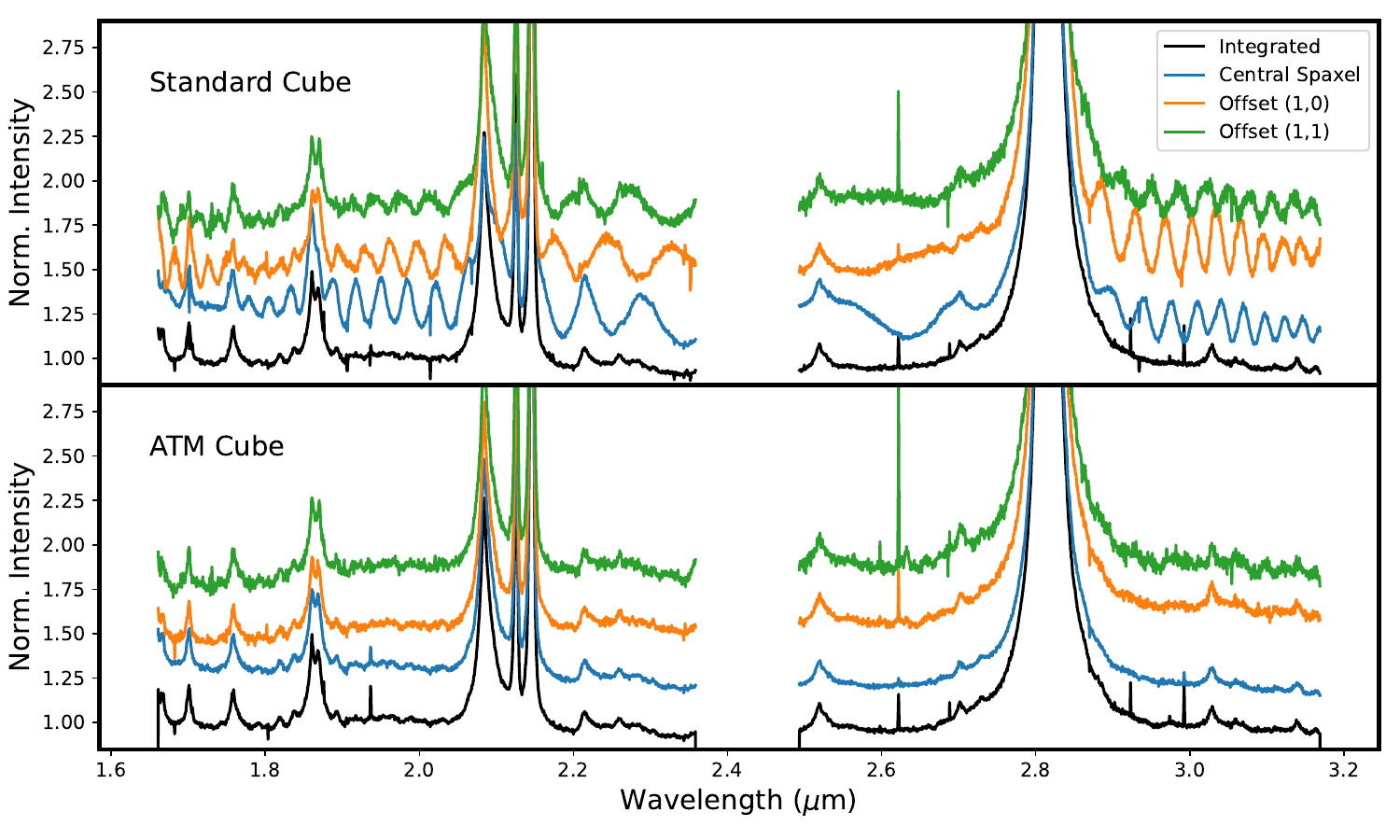}
\caption{Undersampling correction applied to NIRSpec G235H/F170LP observations of a redshift $z \sim 3.3$ QSO from PID 1220.  Top panel: Aperture-summed spectrum (black line) and single-spaxel spectra (colored lines) near the QSO centroid in the standard pipeline data cube.  Bottom panel: As above, but for the adaptive trace model (ATM) oversampled data cube.
}
\label{apt1220.fig}
\end{figure*}

We show the spectrum of LBQS 0302-0019 in Figure \ref{apt1220.fig}.  As illustrated in the top panel (solid black line) the aperture-integrated spectrum shows characteristic broad \Ha\ and \Hb\ emission (redshifted to $\sim$ 2.8 $\micron$ and 2.1 $\micron$ respectively), narrower \othree\ 4959 and 5007 \AA\ forbidden-line emission, and a host of weaker emission lines including He I, H$\delta$, and H$\gamma$.  These emission lines are severely distorted by resampling noise in the spectra of individual spaxels (blue, orange, and green lines), and fainter features can be lost entirely given the large amplitude of the resampling artifacts.
After application of the ATM technique however, these resampling artifacts have been almost entirely eliminated (Figure \ref{apt1220.fig}, lower panel) and fainter spectral features present in the integrated spectrum can be clearly discerned in the individual spaxel spectra.
We provide additional examples of the performance of the ATM algorithm for a variety of different science cases using both the NIRSpec and MIRI IFUs in Appendix \ref{appendix.sec}.

\begin{figure}[!htbp]
\epsscale{1.2}
\plotone{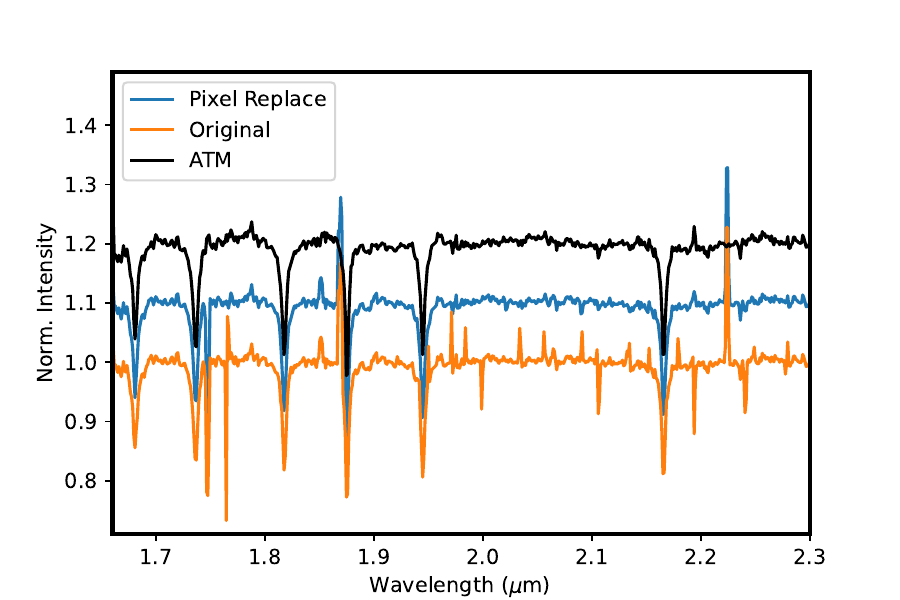}
\caption{Segment of the aperture-summed spectrum from a single exposure of the standard star J17571132 (PID 1536) illustrating the impact of various pipeline steps on outliers in the resulting spectra.
The solid orange line represents the spectrum extracted from a cube using only the `cal' file inputs, while the solid blue line represents the spectrum extracted from a cube using inputs that have been processed through the {\sc pixel\_replace} step.  The solid black line shows the spectrum extracted from a cube using inputs that have been processed through both the {\sc adaptive\_trace\_model} and {\sc pixel\_replace} steps.
All spectra have been normalized by a low-order polynomial fit to the stellar continuum and offset vertically for clarity.
}
\label{spikes.fig}
\end{figure}

An additional benefit of this technique is that it can naturally handle missing data caused by cosmic rays or the ever-growing number of bad pixel artifacts on the JWST detectors,
both of which can produce artifacts in the JWST data cubes.
If a bad pixel lands directly atop the bright central trace of a spectrum for instance, drizzle-type resampling methods \citep[e.g.,][]{fruchter02, law23} fill in the output pixel value using small fractional pixel overlaps from adjacent pixels, resulting in an artificial dip in the spectrum that can be particularly noticeable in cubes built from a small number of exposures.  The JWST calibration pipeline \citep{bushouse25} has pixel-replacement methods designed to mitigate such effects in the 2D calibrated data prior to resampling, as illustrated in Figure \ref{spikes.fig}.  However, such corrections are typically limited to single pixels and cannot account for larger holes resulting from clusters of adjacent bad pixels or saturated central traces of bright objects \citep[although see discussion by][]{vd24}.
By effectively performing a detector-based optimal extraction using a dynamically defined kernel, the spline model technique can do a better job of interpolating across larger detector artifacts and hence reduces such artifacts in the extracted spectra further than the existing pipeline pixel replacement step alone (Figure \ref{spikes.fig}, lower panel).  In the limiting case of an isolated point source, full replacement of the detector pixel values using the spline model can help improve the SNR even further, with improvements $\sim$ 30\% observed for some faint sources.

\begin{figure*}[!htbp]
\epsscale{1.2}
\plotone{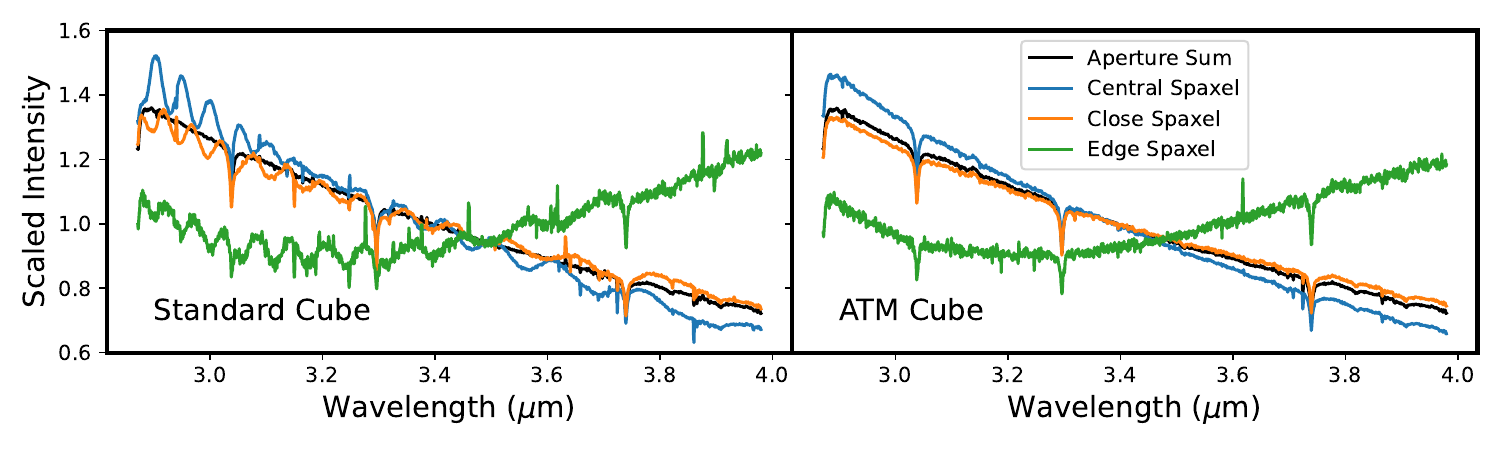}
\caption{Spectra of standard star J1757132 observed with a 16-point dither and the NIRSpec G395H/F290LP grating in standard (left panel) and adaptive trace model (ATM) oversampled (right panel) data cubes, focusing on the NRS1 wavelengths.  The solid black line shows the aperture-integrated spectrum, while blue, orange, and green lines respectively show spectra from the central spaxel, spectra from a nearby spaxel, and spectra from a spaxel far out in the wings of the PSF.  Even after correcting for sampling artifacts, global shape differences between the single-spaxel spectra remain due to the evolving PSF structure as a function of wavelength.
}
\label{nrs3399.fig}
\end{figure*}

We do note, however, that even after application of the ATM technique spaxel-by-spaxel analyses will still have to contend with spectral structure arising from the slow variation in the PSF with wavelength.  Figure \ref{nrs3399.fig} (left-hand panel) for instance shows the aperture-summed spectrum of standard star J1757132 (black line), along with single-spaxel spectra at the center (blue line), slightly off-center (orange line), and near the edge (green line) of the PSF.  The oscillatory artifacts due to resampling noise are well-corrected by the ATM technique (Figure \ref{nrs3399.fig}, right-hand panel) but retain a low-order difference in the slopes of the single-spaxel spectra compared to the composite spectrum.  This effect is mild for the spaxels near the center of the PSF, but extreme for the spaxel in the PSF wings (green line), with the spectrum even appearing to turn up beyond 3.3 $\micron$ due to structure within the PSF shifting into the spaxel at longer wavelengths.  For continuum emission from a centrally-concentrated source this is a well-known effect, and standard analysis tools such as {\sc ppxf} \citep{cappellari17} typically use low-order polynomials to account for it.  In cases where the central source has significant spectral structure however, careful attention to proper PSF modeling for a given scene is required.

Oversampling the fluxes also has implications for the uncertainty of the resulting data cubes as it introduces significant covariance between the oversampled pixel values.  Treating this effect fully would be computationally challenging and time consuming.  To first order, however, it is 
sufficient to oversample the detector error arrays using the linear technique described in \S \ref{linear.sec} so long as the resulting oversampled errors are inflated by a factor of $C$ to account for the covariance.  While strictly incorrect at the oversampled pixel level, this correction ensures that the reported signal-to-noise (SNR) ratios of the resampled data cubes do not depend unphysically on the oversampling factor adopted.  Empirically, we find that $C = 0.23 N + 0.77$ suffices to ensure that the reported SNR of the data cubes is conserved to within about 5\%.


\section{Summary}
\label{summary.sec}

We have presented a technique for mitigating resampling noise in data cubes created from JWST NIRSpec and MIRI IFU observations.  This adaptive trace modeling technique uses basis-spline modeling of the dispersed spectral traces within each IFU slice to map observational data onto a higher-resolution grid prior to resampling, thereby reducing the impact of pixel-phase differences along the dispersed spectral traces.  Since this technique builds fully data-driven models that do not require a priori knowledge of the JWST PSF it can be applied successfully to extremely complicated astronomical scenes.
A key ingredient to the performance of this algorithm in complex scenes is the inclusion of a `residual' correction that accounts for deviations from the spline model in exchange for some loss of SNR relative to what can be achieved used pure spline models for isolated point sources.

This technique has been implemented as an optional step \textsc{adaptive\_trace\_model} in  build 12.3 of the JWST calibration pipeline, and has the potential to significantly aid scientific analyses of JWST IFU observations.

\vskip 20pt

\noindent 
DRL thanks Marshall Perrin for productive conversations and suggestions.
This work is based on observations made with the NASA/ESA/CSA James Webb Space Telescope. The data were obtained from the Mikulski Archive for Space Telescopes at the Space Telescope Science Institute, which is operated by the Association of Universities for Research in Astronomy, Inc., under NASA contract NAS 5-03127 for JWST. These observations are associated with programs \# 1220, \# 1328, \# 1536, \# 1794, \# 1854, \# 1939, and \#3399,  and can be accessed via DOI 10.17909/ek1x-dy66.


\begin{appendix}
\renewcommand\thefigure{\thesection.\arabic{figure}}
\setcounter{figure}{0}   

\section{Applications to Assorted Science Cases}
\label{appendix.sec}

In this appendix, we further illustrate the performance of the ATM technique by applying it to data from a variety of different science cases.

\subsection{MIRI: AGN embedded in a star-forming disk}

A common use case for MIRI MRS is to observe galaxies in which compact AGN are embedded within an extended star-forming disk.  As an example, we take dithered observations of nearby galaxy NGC 7469 from PID 1328 (PI: L. Armus), which contains a Seyfert nucleus surrounded by a star-forming ring \citep{u22}.

\begin{figure*}[!htbp]
\epsscale{1.2}
\plotone{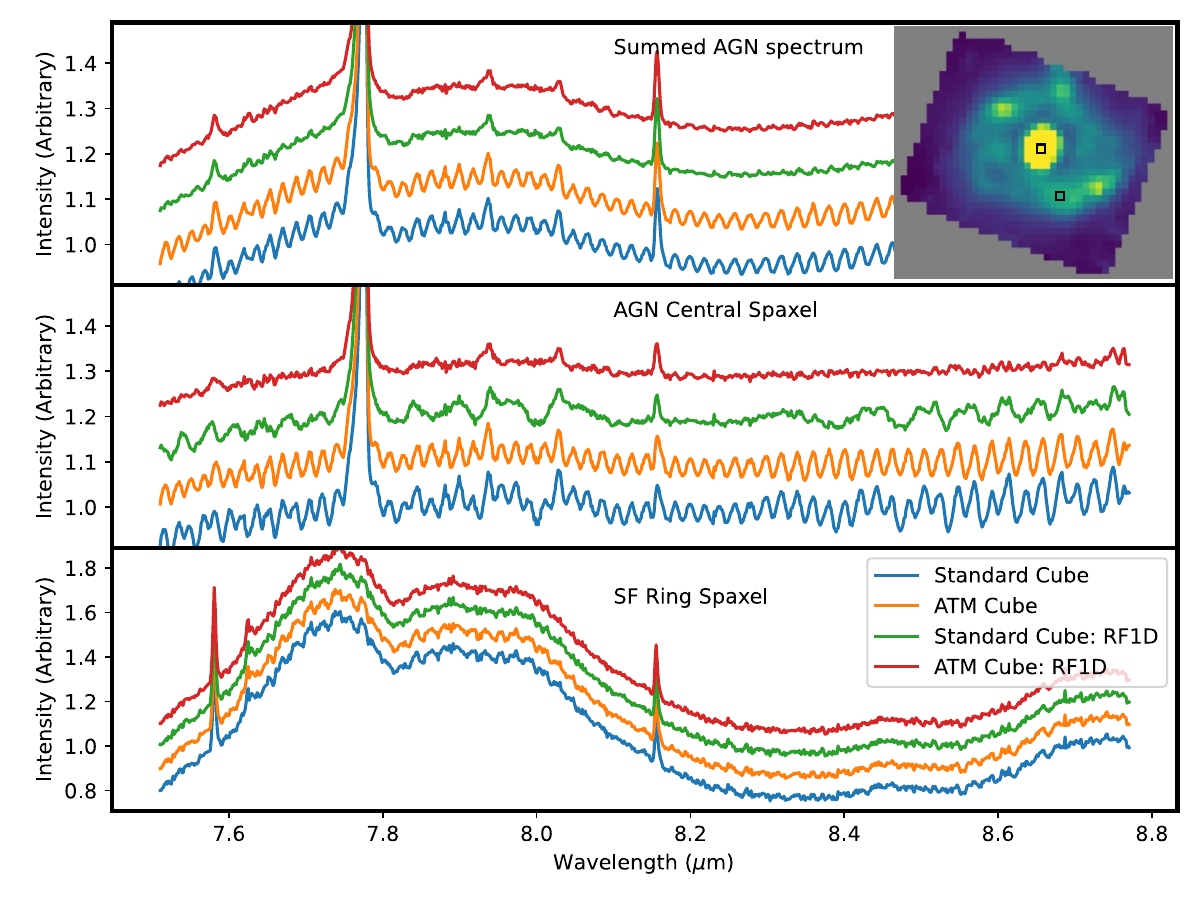}
\caption{MIRI MRS observations of nearby galaxy NGC 7469, with a central Seyfert nucleus surrounded by a star-forming ring.  Top panel: Aperture-summed spectrum of the Seyfert nucleus in the standard vs ATM-corrected data cubes before (blue/orange lines) and after (green/red lines) application of the 1d residual fringe subtraction.  The inset figure shows a continuum image of the source, with black boxes showing the location of spaxels whose spectra are plotted below.  Middle panel: As above, but for the central bright spaxel of the Seyfert nucleus.  Bottom panel: As above, but for a spaxel in the star-forming ring.
}
\label{apt1328.fig}
\end{figure*}

In Figure \ref{apt1328.fig} we show the results of the ATM technique applied to the MRS data.
In the aperture-summed spectrum of the Seyfert nucleus (top panel) we note significant high-frequency sinusoidal structure due to residual fringing; after removing this using the 1d residual fringe correction available in the JWST pipeline we find that the spectra from original and ATM-oversampled data cubes are nearly identical with overall flux calibration that match to better than 0.1\%.
In the central bright spaxel (middle panel), resampling artifacts visible in the original data cube (green line) have been removed in the oversampled data cube (red line), making it easier to see genuine spectral features.
As expected, the oversampling algorithm makes no obvious difference to spectra extracted from the star-forming ring, since resampling noise does not affect such regions of extended emission.

\subsection{NIRSpec: Elliptical galaxy}

While resampling artifacts are most severe in the vicinity of point sources, even extended sources that are sufficiently compact can be affected.  PID 1794 (PI: S. Suyu) for instance observed a quasar that had been gravitationally lensed by a massive elliptical galaxy at redshift $z=0.295$, and \citet{treu25} used spatially-resolved stellar kinematics of the elliptical lensing galaxy to help break the mass-sheet degeneracy.  Efforts to use the PPXF analysis package \citep{cappellari23} to model the Ca triplet absorption feature on a spaxel-by-spaxel basis were significantly complicated by resampling noise, leading \citet{shajib25} to develop their own method for modeling and removing these artifacts from extracted 1d spectra.

\begin{figure*}[!htbp]
\epsscale{1.2}
\plotone{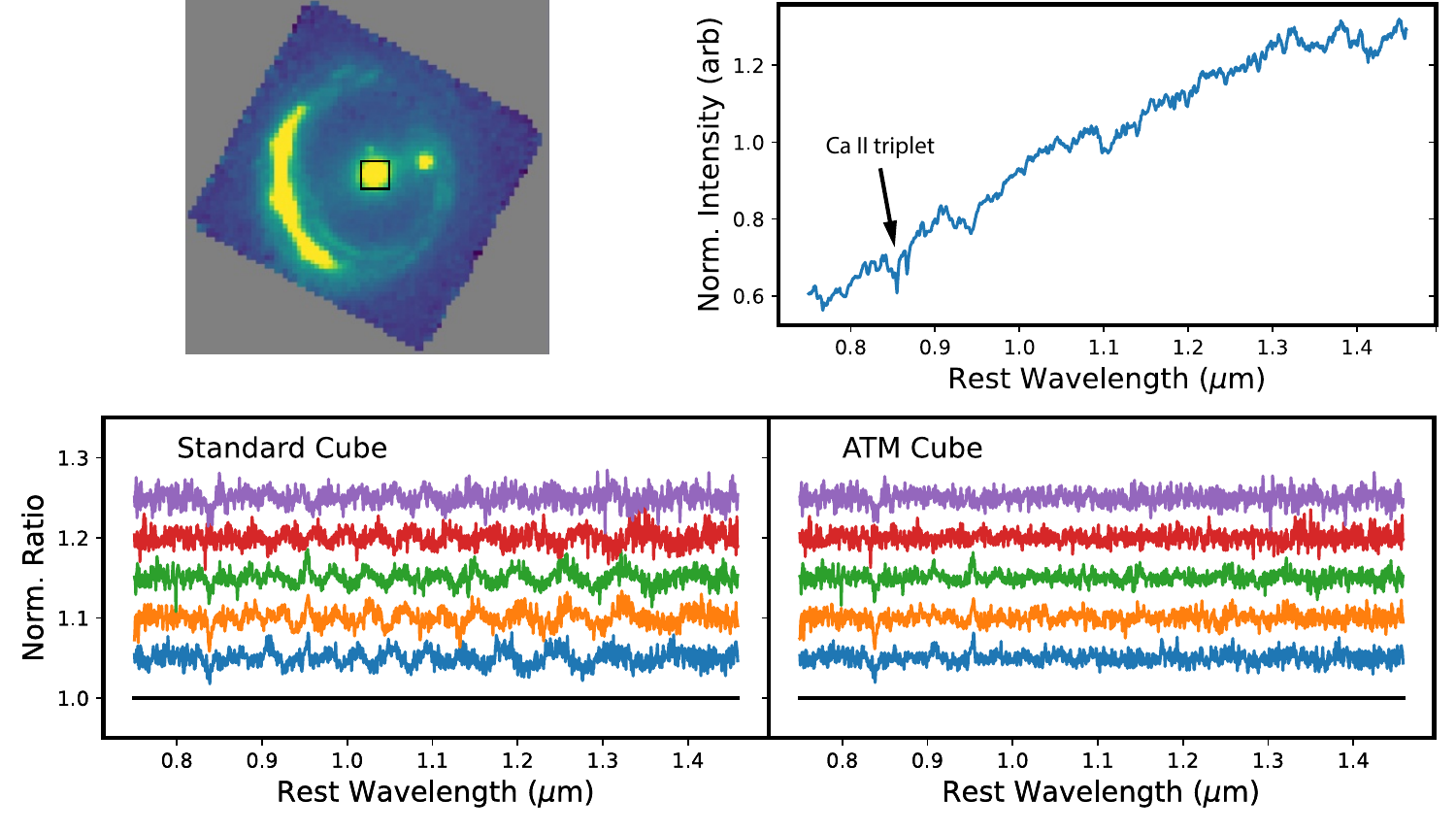}
\caption{Top left panel: continuum image of a NIRSpec IFU observations of an elliptical lensing galaxy from PID 1794.  The black box represents the location of the summed extraction aperture.  Top right panel: Aperture-summed spectrum of the lensing galaxy in rest-frame wavelengths, highlighting the Ca triplet region crucial for kinematic analyses.  Bottom panels: Spectra from the aperture-summed region (solid black line) and five central-most spaxels (colored lines) in the standard and ATM-oversampled data cubes normalized by the aperture-summed spectrum and offset for clarity.
}
\label{apt1794.fig}
\end{figure*}

We illustrate the geometry of this system in Figure \ref{apt1794.fig}, and plot the normalized spectrum extracted from the central elliptical galaxy.  This spectrum has significant structure due to stellar spectral features, most prominently the Ca II absorption triplet.  Since resampling artifacts are thus difficult to see in the spectra of individual spaxels we divide the spectra of these spaxels by the summed galaxy spectrum and normalize the result by a low-order polynomial fit in order to remove slow trends due to the evolution of the PSF FWHM with wavelength.  The resulting normalized spectra are plotted in the lower-left panel of Figure \ref{apt1794.fig}.  The black line in each case is unity by construction, while the colored lines extracted from the standard cube show significant resampling artifacts which can bias the absorption line kinematics.  In contrast, no such artifacts are apparent in similar spectra extracted from the data cube that has been oversampled using the ATM technique, and genuine spatial differences in spectral features such as the Ca II triplet can be more clearly distinguished.

\subsection{NIRSpec: Protoplanetary disk with molecular emission}

\begin{figure*}[!htbp]
\epsscale{1.2}
\plotone{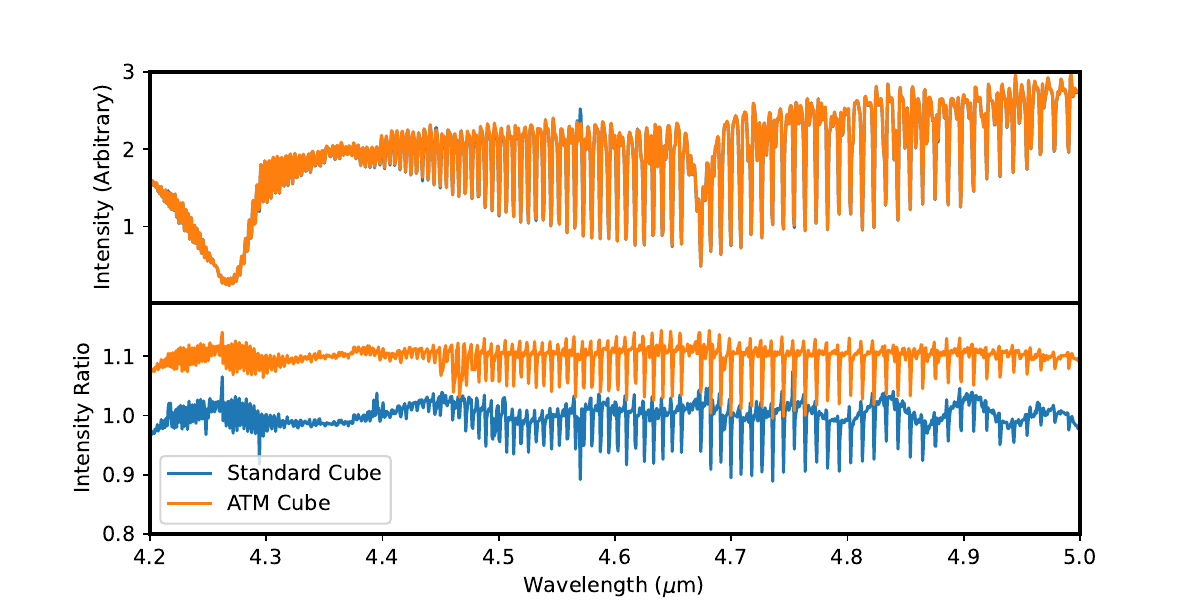}
\caption{Top panel: Aperture-integrated spectra of HH 26-IRS from PID 1854 from both original (solid blue line) and oversampled (solid orange line) data cubes.  Bottom panel: Ratio of the brightest central spaxel to the aperture-summed spectrum for both original and oversampled data cubes.  Ratios have been additionally normalized by a low-order polynomial and shifted vertically for clarity.
}
\label{apt1854.fig}
\end{figure*}

One significant advantage of addressing resampling noise at the detector level, rather than correcting it post-facto in the resulting 1d spectra, is that it can handle spectra of arbitrary complexity.  To illustrate this, we use NIRSpec observations of bright protoplanetary disk HH 26-IRS drawn from PID 1854 (PI: M. McClure).
As illustrated in Figure \ref{apt1854.fig}, the source shows strong and periodic molecular bands \citep[e.g.,][]{mcclure25} that would make it difficult for post-facto techniques to identify and remove the resampling signal and not genuine spectral structure.  The ATM technique produces an aperture-summed spectrum that matches the original to better than 0.3\% (Figure \ref{apt1854.fig}, upper panel), while low-frequency resampling artifacts visible in the central-spaxel spectrum are largely removed by the spline oversampling routine (Figure \ref{apt1854.fig}, lower panel).

\subsection{NIRSpec: Galactic Center}

Finally, we test the ATM method on an extremely complex scene with many overlapping point sources
embedded in diffuse spectral line emission extending throughout the field of view.
We use as our example a 3x3 mosaic of NIRSpec IFU data from PID 1939 (PI: J. Lu) targeting the Milky Way
Galactic center in the G235H/F170LP grating.  This mosaic is undithered (exacerbating the impact of resampling noise), while the crowded sources increase the importance of being able to obtain reliable spectral extractions in extremely small apertures.  In this case the ATM method benefits from ensuring that the spline modeling is applied to all slices regardless of their total flux to ensure that it affects both bright and faint stars.

\begin{figure}[!htbp]
\epsscale{1.2}
\plotone{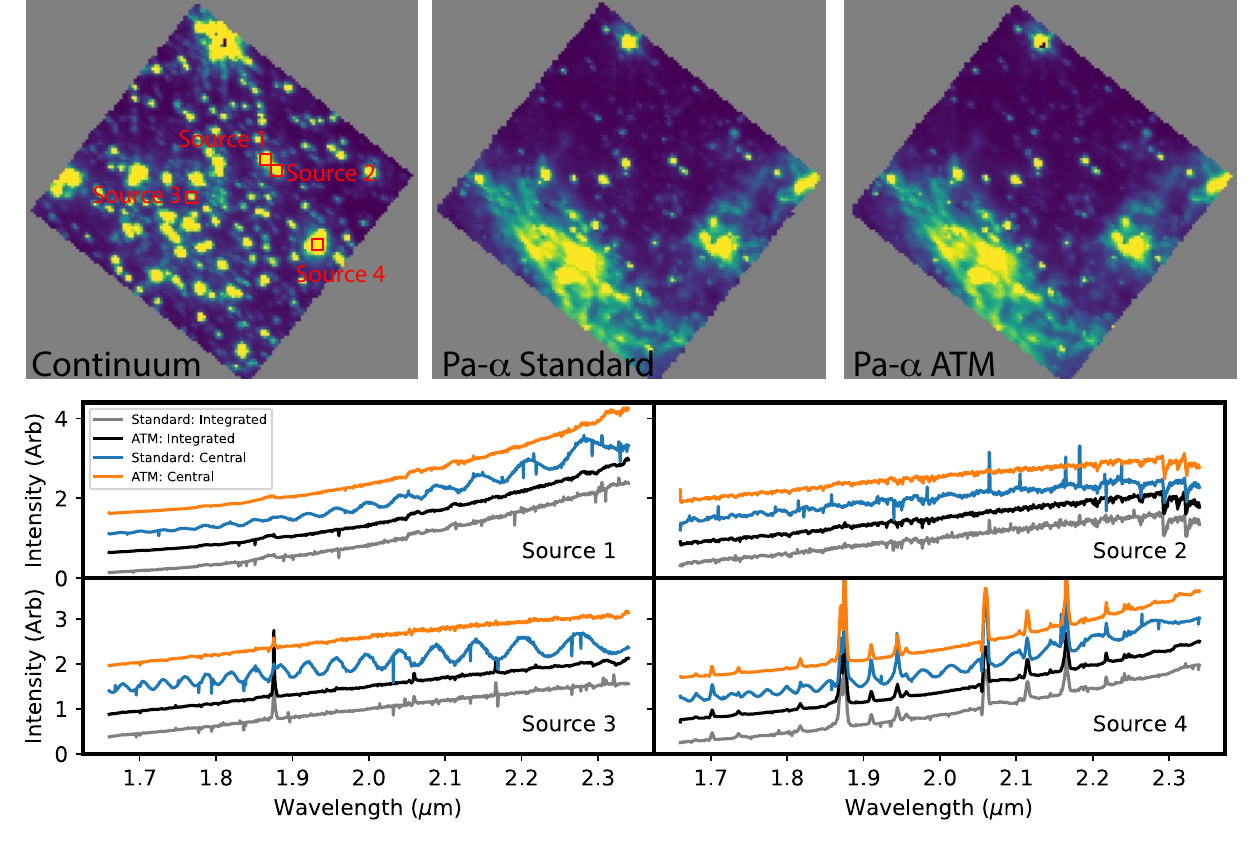}
\caption{Performance of the spline oversampling algorithm in a complex scene from PID 1939 observing the Galactic center using the NIRSpec G235H/F170LP grating.  Top left panel: Image from the mosaicked data cube showing the stellar continuum emission from hundreds of overlapping stars.  Red boxes highlight four stars whose spectra are show below.  
Top middle and top right panels: Standard and ATM-oversampled data cubes at the wavelength of bright Pa-$\alpha$ emission that extends throughout the field of view.
Bottom panel: Spectra of four example sources extracted from the standard and ATM-oversampled data cubes.  Aperture-extracted spectra are shown as grey and black lines, while spectra of the brightest central spaxels are shown as blue and orange lines.  
}
\label{galcent.fig}
\end{figure}

We show the results of the oversampling in Figure \ref{galcent.fig}, focusing on the NRS1 spectral range for clarity.  In all four examples sources the aperture-integrated spectra (grey and black lines) conserve flux to better than 0.4\%, while eliminating some spectral artifacts due to bad and/or missing pixels.  The 
significant resampling artifacts present in the central-spaxel spectra in the standard data cubes (blue lines) have been largely eliminated in the ATM-oversampled data cubes (orange lines) and better match the integrated spectra.
At the same time, we note that the spatial morphology of scene-filling Pa-$\alpha$ emission is unaffected by our spline modeling, thanks to the residual correction term described in \S \ref{residual.sec}.

\end{appendix}

\end{document}